\documentclass[12pt]{article}
\usepackage{epsfig,graphicx}
\begin{document}
\title{The Pressure underneath a Skate at rest}
\author{J. M. J. van Leeuwen}
\maketitle
\begin{center}
Instituut-Lorentz, Universiteit Leiden,\\
Niels Bohrweg 2, 2333 CA Leiden, The Netherlands.
\end{center}
pacs: material science, rheology, deformation of solids.
\begin{abstract}
  The pressure distribution is calculated underneath a skate which is pushed in the ice
  by the weight of a skater at rest. Due to the sharp edges of the skate the
  deformation is partly elastic and partly plastic.
  The ratio of the plastic and elastic contribution to the reaction force is determined. 
  Using this ratio the deformation in ice with a finite hardness can be mapped on the
  problem of the deformation in a purely elastic medium with infinite hardness.
  Both the upright skate and the tilted position are exactly calculated.
\end{abstract}


\section{Introduction}

Skating is centuries old,  but the reason why ice is so slippery that one can skate on it, is
still controversial. Recently several theories \cite{lozowski, pomeau, vanl, vanl2,
weber, smit, canale} have been put forward explaining the low friction of steel on ice.
Indeed it is a quite intricate problem to understand what happens between a skate and
the ice, when a skater moves with a speed of 10 m/s and more. One class of theories
explains the low friction by the formation of a layer of water due to the heat generated
by the friction. Then the crucial problem is to find the pressure distribution in this water
layer. The lubrication approximation yields the solution, but this solution depends on
the assumed boundary conditions at the contact surface.
The other class seeks the explanation in the structure of the surface of ice.
As already suggested by Faraday \cite{faraday}, the surface of ice is wet,
also at rest, i.e. the surface molecules have a high mobility. 
This has been recently confirmed by precision experiments of Weber et al. \cite{weber} 
and Smit et al. \cite{smit}. For this explanation of low friction is it necessary that the 
structure of the surface is  not destroyed or in other words that the deformation is elastic.

So it is important to determine the balance between elastic and plastic deformation.
An indicative answer is given by the pressure distribution 
for a skate at rest, which is a  simpler problem as there is no created water layer
involved. Moreover a perfectly polished skate  cannot exert tangential forces in the
contact surface. The problem of the pressure distribution at rest is a well-defined
problem in the linear elastic approximation. 
The determination of the static pressure distribution is also interesting for other
reasons. Objects with sharp edges, like skates, lead to divergencies in the pressure
distribution. For a skate pressed into the ice we may distinguish three cases
\begin{itemize}
\item The skate has a sufficient large tilt with respect to the normal.
  Then one edge of the skate indents the
  ice and the other edge remains up in the air (see Fig.~\ref{fig1}).
\item The skate has a (very) small tilt, such that also the other edge touches the ice.
  Even for speed skates of small width ($w=1.1$mm), the tilt must be of the order
  of 0.01 radials in order to be in this regime.
\item The skate is perfectly upright. In this case there is symmetry between the
  two edges of the blade.
\end{itemize}
The reason to distinguish these regimes is that each has typical singularities in the
pressures distribution.

The upright skate can be related to the problem of a rectangular stamp, for which
an exact solution exist \cite{musk, johnson}. We can use this solution, but we must
complete it, taking the finite length of the skate into account. The most common
situation is the tilted skate with a sufficient large tilt. For this case we construct
an exact solution. If the tilt angle $\phi=\pi/4$ one can use the exact solution for
a symmetric wedge \cite{john1}, again to be completed to a finite length skate.
The small angle tilt is the most difficult problem for which we did not find an exact
solution. We focus on the first category mentioned and give for information the
complete solution of the perfectly upright skate. This leaves a small gap in our
analysis concerning the small tilts in between. Fortunately this regime is quite
small in practice. 

Our analytic solution takes advantage of the fact that a skate is much longer
than its width, implying that one may, in first approximation, ignore the variations
of the pressure with the length of the skate and later correct for the finite length of the
skate. For an infinite length skate the powerful plane theory of
elastic deformation applies \cite{musk, johnson}.
The plane theory solution
leaves two parameters undetermined, one is related to the depth $d$ of the 
indentation and one is related to the width of the contact zone. 
As the skate is circularly in shape in the longitudinal direction, the depth $d$ can be
linked to contact length $l$ along the skate.
The weight of the skaters provides one restriction on these parameters, the other 
condition has to be found from the proper asymptotic behaviour of the deformation 
far away from the skate.
The plane theory of deformation gives the surface deformation and the pressure
distribution as the real and imaginary part of a function which is analytic in the upper
complex half plane.
Real and imaginary parts of analytic functions are related by equations, which are called 
Kramers-Kronig relations in physics and Plemelj equations in mathematics.
The drawback of the plane theory of deformation is that there is no systematic way to
construct the analytic function. The only way to proof the validity of the
proposed analytic solution is to show that it obeys all the conditions. 

An elastic approximation leading to diverging pressures is, however, internally
inconsistent, unless one assumes an infinite hardness.
The hardness is the maximum pressure
of an elastic deformation. Unfortunately there is a large spread  in the measurements 
of the hardness of ice \cite{lozowski, weber, pourier, penny}.
For our numerical calculations we take the compromise value of the 10 MPa.
Once the elastic
pressure exceeds the hardness, the elastic approximation fails and one must use a 
different rheology. The simple rheology, used in this paper, keeps the 
elastic pressure till the hardness is reached and replaces the elastic pressure 
by the hardness for larger values. 

First we formulate the geometry of the skate. Then we give the elastic deformation
equations and discuss the plane theory in which the variations in the $x$ direction
are ignored. Next we isolate the  two free parameters, for which we derive equations:
one based
on the externally applied force and one following from the asymptotic match with the 
finite skate solution. The details of the analytic solutions for a infinite wedge and an upright skate are given in appendix \ref{wedge} and \ref{upr}.  After completing the 
elastic approximation (with infinite hardness), we construct the solution for a finite hardness. As the tilted skate is more general than the 
upright skate we focus on the tilted skate and use the upright skate for comparison.

The numerical calculations are carried out for standard conditions: a speed skate of
blade width $w=1.1$mm and curvature $R=22$m and a skater of weight 75 kg.

\section{Geometry of the Skate}

\begin{figure}[h]
\begin{center}
  \epsfxsize=0.7\linewidth  \epsffile{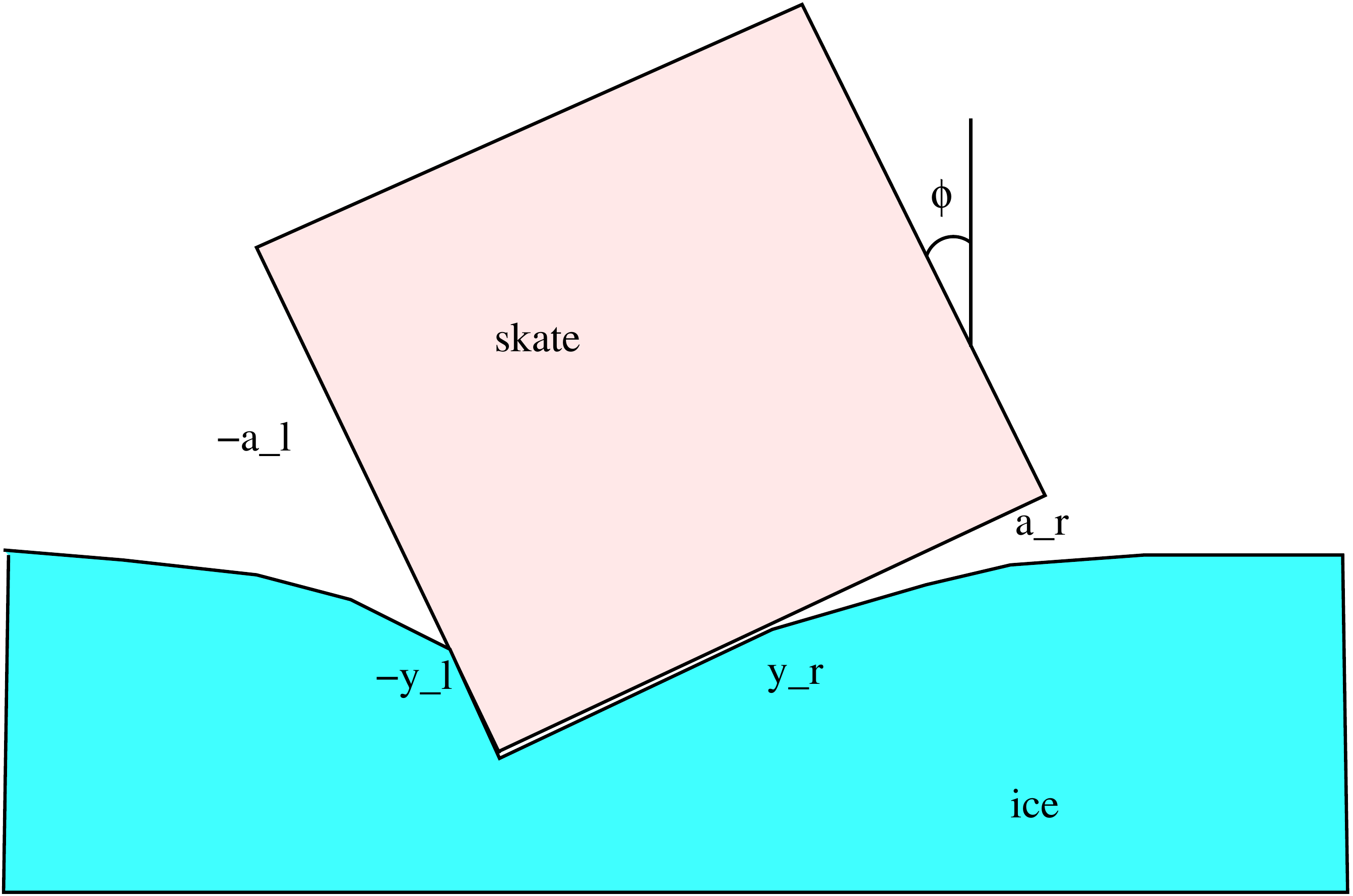}
    \vspace*{0.2cm}

  \caption{Tilted skate indenting the ice. }
\label{fig1}
\end{center}
\end{figure}

A speed skate is almost flat, but due to the curvature it touches the ice over a
contact length $2l$, which is shorter than the length of the skate.
We consider the case where the
skate makes a tilt angle $\phi$ with the normal to the ice. The coordinate system
in which we describe the deformation of the ice, has the $x$ axis in the
longitudinal direction of the skate, the $y$ axis in the transverse direction and the
$z$ axis perpendicular to the undeformed ice.  The ice fills the half space $z<0$.
The origin of the coordinate system is taken in the middle of the skate above
the deepest indentation in the $z$ direction at the level of the (undeformed)
surface of the ice.
The cross-section for a given value of $x$, looks in the $y,z$ plane as shown in
Fig.~\ref{fig1}. The left hand side of the indentation is a surface with the slope
$-a_{\rm l}=-1/\tan \phi $ and the right hand side is a surface with slope
$a_{\rm r}=\tan \phi $.
For skating there is a marked distinction between the two surfaces. The right hand
surface corresponds to the bottom of the blade, which is pushed into the ice with
a large force. The left hand surface is the side of the blade which suffers little force
during motion. For the static skate the two sides are equivalent, each being pushed
into the ice, with a force depending on the tilt angle $\phi$.

The edge of the skate is part of a large circle with radius $R$. The locus $-d(x)$
in the $z$ direction is given by the function
\begin{equation} \label{a1}
  d(x) = \left(d - \frac{x^2}{2 R} \right) \cos \phi,
\end{equation}
approximating the circle by a parabola. The $\cos \phi$ corrects for the tilt.
The contact length $l$ is the point
$x=l$ where the indentation ends. We set
\begin{equation} \label{a2}
  d-\frac{x^2}{2 R} = d [1- (x/l)^2], \quad \quad
  {\rm with} \quad \quad l^2 =2 d R.
\end{equation}
The skate touches the ice in an interval between $-y_{\rm l}$ at the left hand side and
$y_{\rm r}$ at the right hand side.  We call the region $-y_{\rm l} \leq y \leq y_{\rm r}$
the basin of the deformation. 

\section{The Plane Theory relations}

If we may ignore the variation in the $x$ direction with respect to the faster
variations in the $y$ and $z$ the direction, the elastic equations reduce 
to the plane theory for the deformation ${\bf u} (y,z)$ in the $y,z$ plane 
\cite{musk, johnson}.
We have outlined  this reduction in Appendix \ref{elas}
and shown that all quantities can be expressed in the two derivatives of the
$z$ component $u_z(y,z)$ in the $y$ and $z$ direction. Consider the functions
\begin{equation} \label{w1}
  p(y)= 2 \frac{1-\nu}{1-2 \nu}  \left(\frac{\partial u_z(y,z)}{\partial z}\right)_{z=0}
\end{equation}
and
\begin{equation} \label{w2}
  q(y) =\frac{\partial u_z(y,0)}{\partial y}.
\end{equation}
Note that these functions relate to the deformation of the surface $z=0$.
$p(y)$ gives the surface force distribution which is the reaction of the ice on
the indentation. $q(y)$ is the slope of the surface indentation in the $y$ direction. 
The two functions are the real and imaginary part of a complex function $H(y)$,
which is analytic in the upper half of the complex $y$ plane
\begin{equation} \label{w3}
  q(y)= {\rm Re} \, H(y), \quad \quad \quad p(y) =- {\rm Im} \, H(y).
\end{equation}
In Appendix \ref{biharm} we give  the proof of these relation
by an expansion in harmonic functions.
This has the advantage that the whole deformation field
is discussed and that it provides an independent proof for the Kramers-Kronig relations
between the pressure distribution and the surface deformation. However the
Eqns.~(\ref{w3}) show that the problem of the surface indentation is self-contained.

Eq.~(\ref{w3}) is equivalent with the Kramers-Kronig relations which express
$p(y)$ in terms of $q(y)$ and vice versa.
Since there is a multitude of complex functions, which are analytic in the
upper half $y$ plane, the specific solution has to be  determined
from the boundary conditions on the real $y$ axis.

\section{The boundary conditions for $q(y)$ and $p(y)$}

The boundary conditions on $q(y)$ and $p(y)$ are of the mixed type. 
$q(y)$ is dictated inside the basin by the shape of the skate and is therefore known
inside the basin. 
Outside the basin $q(y)$ is unknown, while $p(y)$ is unknown
inside the basin and vanishes outside. The inside value of $q(y)$ follows from
$u_z(y,0)$ as
\begin{equation} \label{c1}
  \left\{ \begin{array}{rclc}
            u_z(y,0) &= &-d - a_{\rm l} y, \quad \quad& -y_{\rm l}  < y < 0,\\*[1mm]
            u_z(y,0) &= &-d + a_{\rm r} y, \quad \quad &0 < y < y_{\rm r},
          \end{array} \right.                                                                                             
\end{equation}
where $a_{\rm r}=\tan \phi $ viz. $a_{\rm l}= -1/\tan \phi $ are the right viz. left
slope of the wedge.
We call $a_{\rm r}$ the tilt slope, it varies from $a_r=0$ (upright) to $a_r=1$.
Larger tilts are unrealistic and could be treated by interchanging left and right.
Differentiation of the Eqns~(\ref{c1}) with respect to $y$ gives the value of $q(y)$
inside the basin. Thus we arrive at the following boundary conditions.
\begin{equation} \label{c2}
  \left\{ \begin{array}{rcll}
            p(y) &= & 0, &  \quad   y < -y_{\rm l} , \\*[2mm]
            q(y) & = & -a_{\rm l}, & \quad -y_{\rm l} < y < 0,  \\*[2mm]                      
            q(y) &= & a_{\rm r}, &  \quad 0 < y < y_{\rm r},\\*[2mm]
            p(y) & = & 0, & \quad  y > y_{\rm r}. \\*[2mm]
          \end{array} \right.
\end{equation}
We see that $q(y)$ has a jump $A=a_{\rm l} +a_{\rm r}$ at the origin $y=0$ and 
that $p(y)$ changes from
a non-vanishing function to zero at the boundaries of the basin. These are the
singularities that dictate the structure of the analytic function of which $q(y)$ and
$p(y)$ are the real and imaginary parts.
Frequently occurring combinations of the boundaries $-y_{\rm l}$ and $y_{\rm r}$,
are the arithmetic mean $m$ and the ratio $r^2$
\begin{equation} \label{c3}
  m= \sqrt{y_{\rm r} y_{\rm l}}, \quad \quad \quad r = \sqrt{y_{\rm r}/y_{\rm l}}.
\end{equation}
Inversely we find  $y_{\rm l}$ and $y_{\rm r}$ as
\begin{equation} \label{c4}
y_{\rm l} =m /r, \quad \quad \quad y_{\rm r} = m r.
\end{equation}
It will turn out that the ratio $r$ is determined by the solution, but that the mean
$m$ is a free parameter in the solution. 

\section{The analytic solution}

In Appendix \ref{wedge} we give the solution for an infinitely long wedge which
has a slope $a_{\rm r}$ in the positive direction and a slope $-a_{\rm l}$ in the
negative direction. The skate is a special case as the edge is rectangular,
implying $a_{\rm r}=\tan \phi$ and $a_{\rm l}=1/\tan \phi$. The generator of the
solution is a function $H(y) $, which is analytic in the upper half complex $y$ plane.
The connection of the functions $q(y)$ and $p (y)$ with $H(y)$ has been given in
Eq.~(\ref{w3}). Here we copy $H(y)$ from Appendix \ref{wedge}.
\begin{equation} \label{d2}
H(y) = a_{\rm r} + \frac{A}{2 \pi i} \log \left( \frac{X(y)+\sqrt{Y(y)}}
    {X(y)-\sqrt{Y(y)}} \right), \quad \quad 0 < y < y_{\rm r}.
\end{equation}
The parameter $A$ and the functions $X(y)$ and $Y(y)$ read
in Eq.~(\ref{c2})
\begin{equation} \label{d3}
  A=a_{\rm r}+a_{\rm l}, \quad \quad  X(y) = m+ R_- y/2, \quad \quad
  Y(s) = (y_{\rm r} - y ) (y_{\rm l} + y),
\end{equation}
with $m$ and $r$ defined in Eq.~(\ref{c3}) and with
$R_{\pm}$ for the combinations
\begin{equation} \label{d4}
R_\pm = r \pm 1/r.
\end{equation} 

The function $H(y)$ is represented by different expressions in the various
regions along the real $y$ axis (see Appendix \ref{wedge}). 
For $p(y)$ the imaginary part inside the basin is important. It reads
\begin{equation} \label{d5}
-{\rm Im}\,  H(y) = \frac{A}{2 \pi} \log\left( \frac{X(y)+\sqrt{Y(y)}}
    {X(y)-\sqrt{Y(y)}} \right),  \quad \quad -y_{\rm r}  < y < y_{\rm r}.
\end{equation} 
For $q(y)$ the real part outside the basin is relevant. It is given by
\begin{equation} \label{d6}
\left\{ \begin{array}{rccl}
          {\rm Re} \, H(y) & = & \displaystyle  a_{\rm r} - \frac{A}{\pi}  
                                 \arctan (\sqrt{-Y(y)}/X(y))
          & \quad \quad y > y_{\rm r} , \\*[3mm] 
          {\rm Re} \, H(y) & =&\displaystyle  -a_{\rm l} + \frac{A}{\pi}  
                                \arctan (\sqrt{-Y(y)}/X(y)) & \quad \quad y <-y_{\rm l}
\end{array} \right.
\end{equation}
From the requirement that $H(y)$ remains well behaved at large $y$ follows the value
of the ratio $r$. It is given in terms of the angle $\chi$ 
\begin{equation} \label{d7}
  \chi = \frac{\pi a_{\rm r}}{A} = \pi \sin^2 \phi.
\end{equation}
The second equality holds for the skate as rectangular wedge. In Appendix
\ref{wedge} it is found that (Eq.~(\ref{A15}))
\begin{equation} \label{d8}
r = \frac{1}{\tan(\chi/2)}.
\end{equation}
While $r$ is thus fixed by the tilt, $m$ remains undetermined by the solution. In fact
the mean $m$ depends on the longitudinal coordinate $x$, as we shall see. 

\begin{figure}[h]
\begin{center}
  \epsfxsize=0.9\linewidth  \epsffile{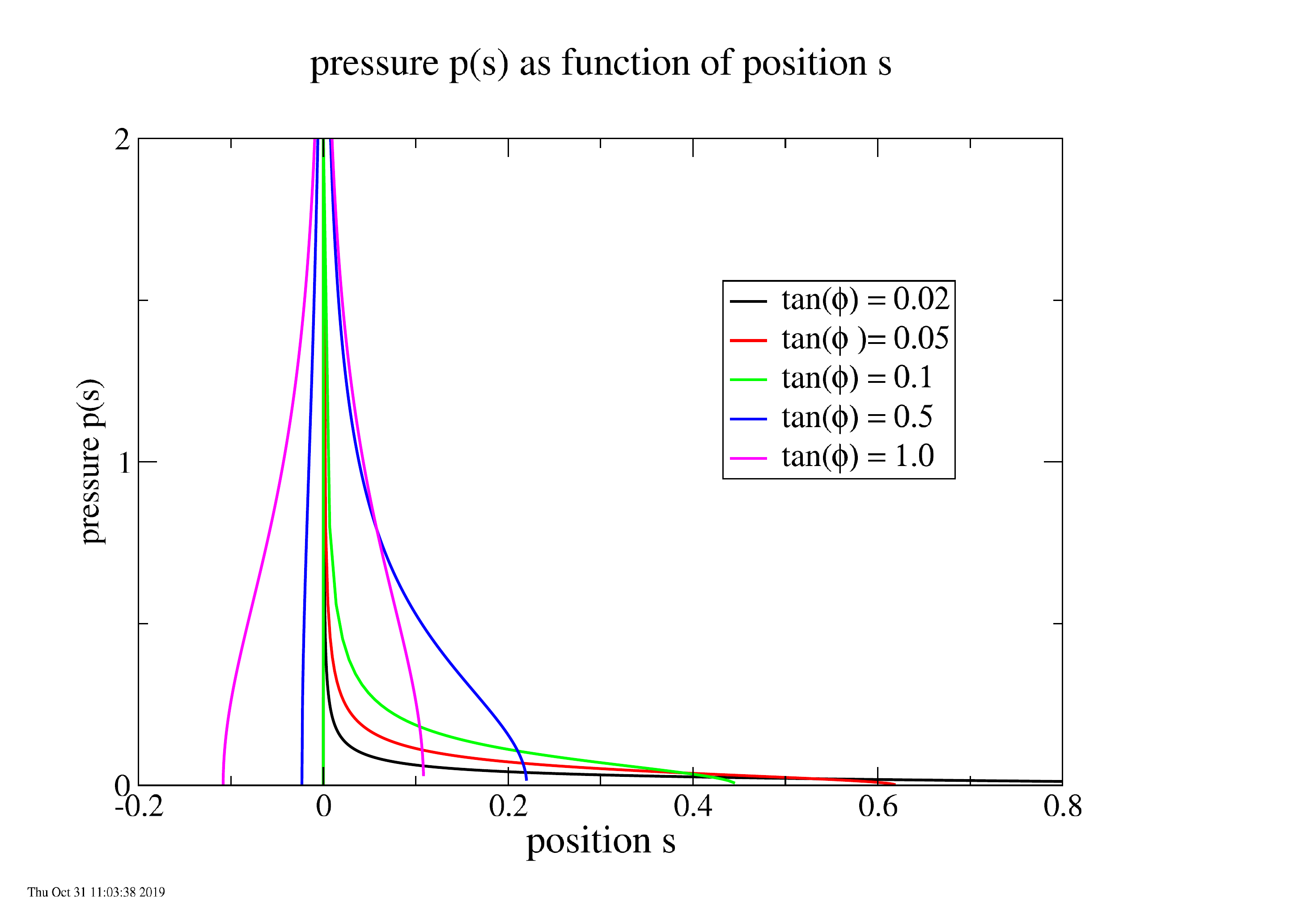}
    \vspace*{0.2cm}

  \caption{p(s) as function of position s=y/m.}
\label{pressure}
\end{center}
\end{figure}

In order to get an idea on the shape of the pressure at the
surface of the ice we have plotted in Fig. \ref{pressure} the function
$p(y)$ for various values of the tilt $a_{\rm r}$. Instead of using the argument $y$,
we employ $s=y/m$ which make the curves independent of $m$.
All curves show a logarithmic singularity at $y=0$. While for
high slopes the range is short, it increases towards the upright position and
would diverge for $\phi=0$ if the description of the tilted skate with one edge
would hold all the way down till $\phi=0$. 

As the function $q(y)$ decays for large $y$ as $1/y$, the function $u_z(y,0)$,
which is the integral of $q(y)$, grows logarithmically, implying  that the 
deformation of the surface keeps growing indefinitely far away from the skate.
This is the result of treating
the skate as infinitely long. Far away from the skate, for $y$ of the order of the
contact length, the solution does not apply and a cross-over to the
real behaviour has to be found.

\section{Elastic Forces on the surfaces}

The determination $m(x)$ turns out to be delicate. The
first ingredient is the calculation of the forces on the contact surfaces. $p(y)$ gives
the force perpendicular to the surface. The total forces are given as the integrals
\begin{equation} \label{e1}
  F_{\rm l} = \frac{E^*}{2} \int_{-l}^l dx \int^0_{-y_{\rm l}} p(y) \frac{dy}{\sin \phi},
  \quad   \quad \quad F_{\rm r} = \frac{E^*}{2} \int_{-l}^l dx \int^{y_{\rm r}}_0
  p(y) \frac{dy}{\cos \phi}.
\end{equation}
The differential $dy$ has to be corrected for the tilt of the interface, as we have to
integrate over the surface and not over the horizontal parameter $y$.
The $\sin \phi$ and $\cos \phi$ in the denominator of the $dy$ element take this
into account. The integration over $y$ can be evaluated, since the primitive of
$p(y)$ is explicitly known (see Eq.~(\ref{A20}). 
\begin{equation} \label{e2}
  \left\{ \begin{array}{rcrcl}
             h_{\rm l} (x) & = & \displaystyle \int^0_{-y_{\rm l}} p(y) dy & = & \displaystyle
                                                \frac{Am(x)}{\pi} \left(\frac{\pi}{2}
                                                -\arcsin \frac{R_-}{R_+} \right),\\*[4mm]
             h_{\rm r} (x) & = & \displaystyle \int^{y_{\rm r}}_0 p(y) dy  &= &\displaystyle
                                               \frac{Am(x)}{\pi} \left(\frac{\pi}{2}
                                               +\arcsin \frac{R_-}{R_+} \right).
          \end{array} \right.
\end{equation}
These expressions can be simplified with Eq.~(\ref{A17}) 
\begin{equation} \label{e3}
  \arcsin \frac{R_-}{R_+} = \arcsin ( \cos \chi)  = \pi/2 - \chi,
\end{equation}
with the result for the forces
\begin{equation} \label{e4}
  \left\{ \begin{array}{rllcl}
            h_{\rm l} (x)  & = & A m(x) (\chi /\pi) & = & A m(x) \sin^2 (\phi), \\*[3mm]
            h_{\rm r} (x) & = & A m(x) (\pi - \chi ) / \pi &= & A m(x) \cos^2 (\phi).
     \end{array} \right.                                                         
\end{equation} 
For the total force on the surface we have to multiply by the integral over $x$
\begin{equation} \label{e5}
  S = \int^l_{-l} m(x) dx.
\end{equation}
$S$ is the (mean) contact surface. Thus we find for the total forces
\begin{equation} \label{e6}
  F_{\rm l} = (E^*A S/2) \sin \phi, \quad \quad {\rm and} \quad \quad
  F_{\rm r} = (E^*A S/2)  \cos \phi.
\end{equation}
For the rectangular skate we may use the $\phi$ dependence of $A$ reading
\begin{equation} \label{e7}
  A = \tan \phi + 1/\tan \phi = \frac{1}{\sin \phi \cos \phi}
\end{equation} 

With the partial forces we can construct the total force. The component perpendicular
to the ice is given as
\begin{equation} \label{e8}
  F_z = F_{\rm l}\sin \phi  + F_{\rm r} \cos \phi =\frac{E^* S A}{2}
\end{equation}
and the component tangential to the ice reads
\begin{equation} \label{e9}
  F_y = F_{\rm l} \cos \phi  - F_{\rm r} \sin \phi = 0
\end{equation}
The calculated $F_z$ a $F_y$ are the forces that the ice exerts on the skate. They have
to be compensated by the total force of the skate on the ice, which thus is purely
in the vertical direction. Therefore we may assume that $F_z$ is given. It determines
the value of $S$, which gives a condition on $m(x)$.

\section{Determination of $m(x)$}

Since the skate is finite, it may be considered, far away from the skate,
as a distribution of point forces. As all the relations refer to the surface $z=0$
we omit the $z$ as an argument. The asymptotic deformation of a point force is
well known and decays inversely with the distance. For the distribution of the
point forces we may take for $f_z(x,y)$
\begin{equation} \label{f1}
  f_z (x,y) = \frac{E^*}{2} p(y) \quad \quad -y_{\rm l} \leq y \leq y_{\rm r}
\end{equation} 
So $f_z$ is known up to the scale $m(x)$ (implicitly in $p(y)$). The surface 
deformation due this $f_z (x,y)$ is given by \cite{landau}
\begin{equation} \label{f2}
  u_z(x,y)  \simeq - \frac{1}{\pi E^*} \int^l_{-l} dx' \int^{y_{\rm r} (x')}_{-y_{\rm l} (x')}
  dy' \frac{f_z(x',y') }{[(x-x')^2+(y-y')^2]^{1/2}}.
\end{equation}
The minus sign accounts for the fact that we have to take the pressure of the
skate on the ice. It is opposite to the pressure in the ice, which we are calculating.
This expression has to match the solution as derived from  $q(y)$ as
\begin{equation} \label{f3}
u_z (x,y) = - d(x) + \int_0^y dy \,  q(y).
\end{equation} 
The matching region is $m(x) \ll y \ll l$, for which Eq.~(\ref{f2}) still holds 
since $m(x) \ll y $ and Eq.~(\ref{f2}) still holds since $y \ll l$. The matching will
imply an equation for $m(x)$. As a result the only remaining adaptable parameter
is the intrusion depth $d$ or equivalently the contact length $l$, which must be 
chosen such that $F_z$ in Eq.~(\ref{f8}) is matched by the external force.

\subsection{Asymptotics far from the center} 

For $y \gg m(x) $ one may drop the $y'$ in the denominator
of Eq.~(\ref{f2}) and use the integral over $p(y)$, given in Eq.~(\ref{A21}).
This turns Eq.~(\ref{f2}) into 
\begin{equation} \label{f5}
    u_z(x,y) \simeq  -\frac{A}{2 \pi} \int^l_{-l} dx' \frac{m(x') }{[(x'-x)^2+y^2]^{1/2}}.
  \end{equation}
The integration has a vanishing denominator for $x'=x$ and $y=0$, which causes
the logarithmic increase for $y \ll l$. The function $m(x')$ is
regular at $x'=x$. So we may expand $m(x')$ around $x'=x$ and only the leading 
term influences the amplitude of the logarithmic increase. The higher orders mask
the singular denominator. So we may write Eq.~(\ref{f5}) as
\begin{equation} \label{f6}
  u_z(x,y)  \simeq - \frac{A m(x)}{2 \pi} \int^l_{-l} dx'   \frac{1 }{[(x'-x)^2+y^2]^{1/2}}.
\end{equation} 
The integration over $x'$ is now elementary, yielding
\begin{equation} \label{f7}
u_z(x,y)  \simeq - \frac{A m (x)}{2 \pi} [{\rm asinh}((l-x)/y)-
{\rm asinh}((-l-x)/y)].
\end{equation}
For $y \ll l$, where one starts to see the skate as very long, the
deformation crosses over to
\begin{equation} \label{f8}
u_z(x,y)  \simeq -\frac{A m(x) }{2 \pi} \log(4(l^2-x^2)/y^2) 
\end{equation} 

\subsection{The asymptotic behaviour  of $u_z$}

For the comparison of the behaviour of $u_z (x,y)$, as given by Eq.~(\ref{f3}),
with that given by Eq.~(\ref{f8}), we have to carry out the integral
\begin{equation} \label{f9}
 \int^y q(y') dy'= \int^y {\rm Re} H(y') dy'= y H(y) +  \frac{A m(x)}{\pi} 
{ \rm acosh} \left( \frac{2 y -y_{\rm r} + y_{\rm l}}{y_{\rm r} + y_{\rm l}} \right),
\end{equation}
which has been worked out in Eq.~(\ref{A23}) in Appendix \ref{wedge}. 
The term $y H(y)$ approaches a constant for large $y$ (see Eq.~(\ref{A25}))
\begin{equation} \label{f11}
  y H(y)_{y \rightarrow \infty} = \frac{A m(x)}{\pi}.
\end{equation}
The other contribution in Eq.~(\ref{f9}) gives asymptotically a logarithmically
increasing term and constant. Dropping the decaying terms we find
\begin{equation} \label{f12}
 u_z (x,y)_{y \rightarrow \infty}  \simeq  -d(x) +\frac{A m(x)}{\pi} 
\left(1+  \log\left(\frac{4 y}{m(x)R_+}\right)\right).
\end{equation}

\subsection{The asymptotic condition}

In Eqns.~(\ref{f8}) and (\ref{f12}) we have two expressions for the asymptotic
behaviour of the surface for a region $m \ll y \ll l$ which both contain an
increasing logarithm and a constant. The coefficients of the logarithm are in
both cases equal to $A m(x)/\pi$. So the expression agree in shape.
In order that the expressions coincide, we must require the constants to be equal
\begin{equation} \label{f13}
  - \frac{A m(x) }{2\pi} \log (4(l^2-x^2)/y^2) = -d(x) +
  \frac{A m(x)}{\pi} (1+  \log[4/(m(x) R_+)]),
\end{equation}
which can also be written as
\begin{equation} \label{f11}
  d(x)= \frac{A m(x)}{\pi} \left[1+
    \log \left(\frac{8 \sqrt{l^2-x^2}}{ m(x) R_+} \right) \right].
\end{equation}
This is the desired second relation between $d(x)$ and $m(x)$.  In first approximation
$m(x)$ is proportional to $d(x)$. The expression between square brackets
implies only a weak $x$ dependence.
\begin{figure}[h]
\begin{center}
  \epsfxsize=0.7\linewidth  \epsffile{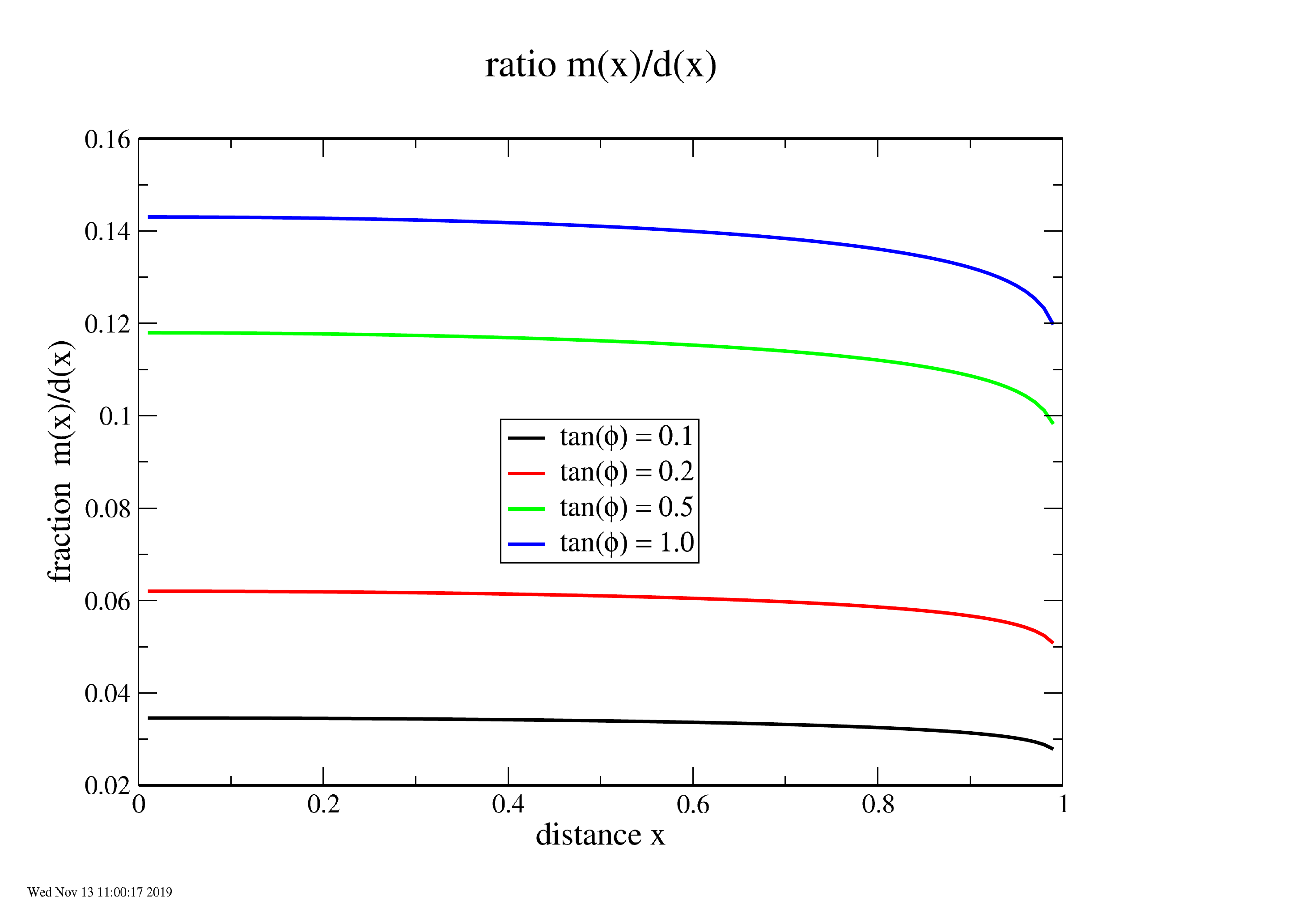}
    \vspace*{0.2cm}

  \caption{mean width $m(x)/d(x)$ as function of $x$. }
\label{mean}
\end{center}
\end{figure}

 In Fig.~(\ref{mean})
we have plotted the function $m(x)/d(x)$ for a number of values of $\tan \phi$. 
As one sees the value is rather constant over the values
of $x$. Only near the tip of the skate $x \simeq l$ the value decreases noticeable.
This effect is not very relevant since $m(x)$ always occurs in combination of $d(x)$,
which goes faster to 0 than $m(x)$ for $x \rightarrow l$. Therefore one does not make
a large error by replacing $m(x)$ by its value $m(0)$ at $x=0$.

\begin{figure}[h]
\begin{center}
  \epsfxsize=0.7\linewidth  \epsffile{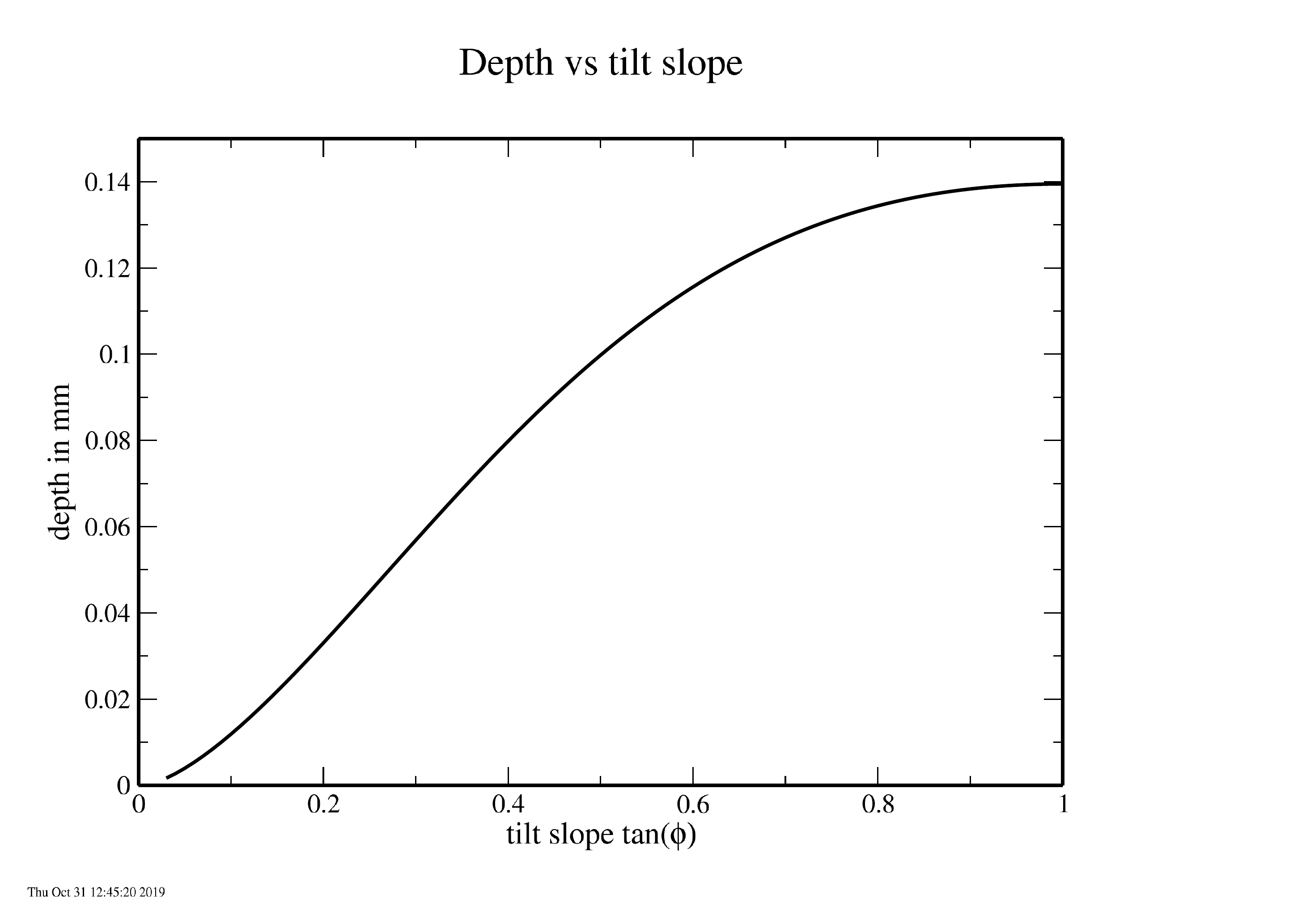}
    \vspace*{0.2cm}

  \caption{Depth $d$ as function of the tilt slope $\tan(\phi)$. }
\label{depth}
\end{center}
\end{figure}

The value of $m(x)$ decreases with the tilt angle $\phi$. At fixed force $F_z$ the
contact area should stay more or less constant. That implies that, if $m(x)$ shrinks,
the contact length $l$ and therefore also the intrusion depth $d$ have to grow.
The depth $d$ has been plotted in Fig.~\ref{depth} as function of the
tilt slope $\tan(\phi)$.
The depth is maximal at large tilt angles. For the upright position it has a value
for the same parameters of hardness and weight, which is quite far below the
calculated values at non-zero tilt angles.

\section{Finite hardness}

In the preceding sections we have assumed that the deformation is purely elastic,
while mentioning at the same time that at the sharp edges the pressure diverges. Elastic
pressures are limited by the hardness of the deformed medium. So the elastic theory
applies all the way if the hardness is infinite. The hardness of ice $p_{\rm h}$ is of 
the order of 10 MPa,  with a large variation in the measured values
\cite{pourier, weber}. When the normal pressure exceeds the hardness,  i.e when
\begin{equation} \label{h1}
E^* p(y) /2 > p_{\rm h}, \quad \quad {\rm or} \quad \quad p(y) > 2 p_{\rm h}/E^*
\end{equation} 
the elastic deformation will turn into a plastic deformation. Since 
$2 p_{\rm h} /E^* \simeq 0.0213$, one observes from Fig.~\ref{pressure} that only
near the boundaries $-y_{\rm l}$ and $y_{\rm r}$ the pressure stays below the hardness.

In general, since the skate meets a lesser resistance than the calculated elastic force,
the skate will intrude deeper into the ice due to the plastic deformation.
A plastic deformation manifests itself when the skate is lifted from the ice. 
A permanent indentation is left behind. In fact from the
size of this indentation and the applied external force on the skate, the hardness of ice
can be measured (as in Brinell hardness measurements). The simple assumption is
that the surface area of the indentation times the hardness equals the applied force.
However, from the pressure distribution we see that not the whole contact area is
plastically deformed by the skate.

In this section we make an estimate of the ratio between elastic and plastic deformation
by replacing the elastic counter pressure by the hardness where it exceeds the hardness. Thus we have to determine the points $-y_{\rm cl}$ and $y_{\rm cr}$
where Eq.~(\ref{h1})
turns into an equality. These points are the solution of the equation
\begin{equation} \label{h2}
p_{\rm h} = E^* \frac{A}{2 \pi}  \log 
  \left( \frac{X(y_{\rm c}) + \sqrt{Y(y_{\rm c})}}{X(y_{\rm c}) - \sqrt{Y(y_{\rm c}}} \right).
\end{equation}	
This equation becomes more transparent by introducing the variable $\alpha$
\begin{equation} \label{h3}
\alpha = \frac{2 \pi p_{\rm h} }{ E^* A},
\end{equation}
which leads to the quadratic equation for $y_{\rm c}$
\begin{equation} \label{h5}
Y(y_{\rm c}) = X^2 (y_{\rm c}) \tanh^2 \alpha.
\end{equation}
The two roots will be denoted by $y_{\rm cr} $ and $-y_{\rm cl} $.
After some algebra one finds
\begin{equation} \label{h6}
y_{\rm cl}= \frac{2 m}{R_+ \cosh \alpha + R_- }, \quad \quad \quad 
y_{\rm cr} = \frac{2 m }{R_+ \cosh \alpha - R_- }.
\end{equation}

Inside the region $-y_{\rm cl} < y < y_{\rm cr}$ the pressure equals $p_{\rm h}$ and 
outside this region the elastic expression for the pressure holds. We can again 
compute the integrals $h_{\rm l} (x) $ and $h_{\rm r} (x)$ as in Eq.~(\ref{e2}). The
integrals are elementary with the result
\begin{equation} \label{h7}
  h_{\rm l} (x) = \int_{-y_{\rm cl}}^0 dy \, p(y) =
  \frac{A m(x)}{\pi} (\pi/2 - \arcsin \beta_{\rm l}) 
\end{equation} 
and 
\begin{equation} \label{h8}\
  h_{\rm r} (x) =  \int_0^{y_{\rm cr} } dy \, p(y) =
  \frac{A m(x)}{\pi}(\pi/2 -\arcsin \beta_{\rm r}) . 
\end{equation}
The values $\beta_{\rm l}$ and $\beta_{\rm r}$ are given by
\begin{equation} \label{h9}
\beta_{\rm l} =\frac{R_+ +R_- \cosh \alpha}{R_+ \cosh \alpha +R_-}, \quad \quad 
\quad \beta_{\rm r} = \frac{R_+- R_- \cosh \alpha}{R_+ \cosh \alpha - R_-}.
\end{equation}

For infinite hardness $\alpha = \infty $, the result coincides with Eq.~(\ref{e2}) 
with $\beta_{\rm l} = - \beta_{\rm r} = R_-/R_+= \cos \chi$.  
For zero hardness $\alpha = 0$, $\beta_{\rm l} = \beta_{\rm r} =1$.
In that case both $h_{\rm l} (x)$ and $h_{\rm r} (x)$ vanish since the ice gives 
no resistance against indentation.

\begin{figure}[h]
\begin{center}
  \epsfxsize=0.7\linewidth  \epsffile{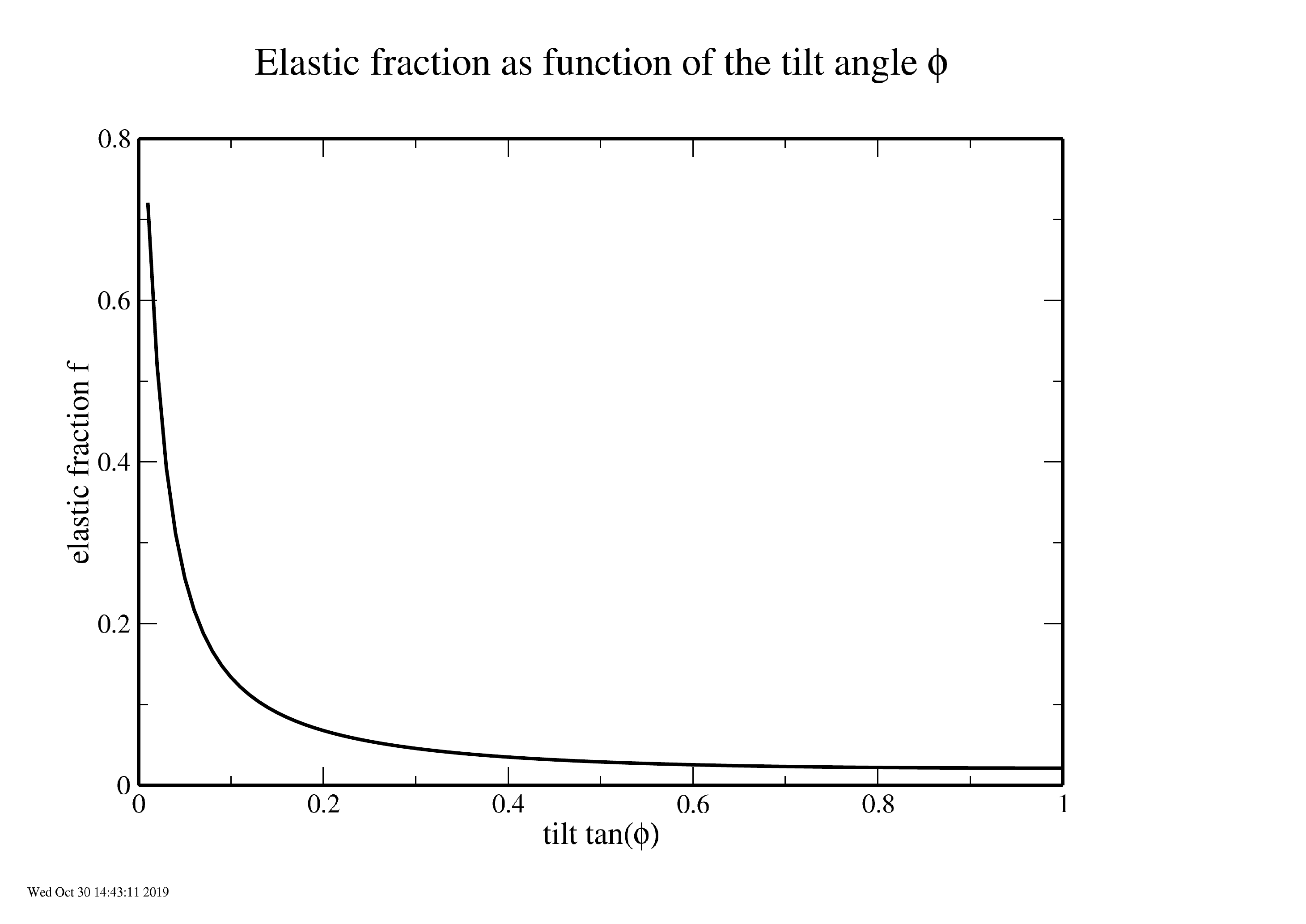}
    \vspace*{0.2cm}

  \caption{Elasticity $f$ versus tilt slope $\tan(\phi)$. }
\label{fraction}
\end{center}
\end{figure}

The external force $F_z$ in case of partial plastic deformation is less than the fully 
elastic counter force $F_{\rm el}$ with the ratio
\begin{equation} \label{h11}
  \frac{F_z}{F_{\rm el}} = f =  (\pi - \arcsin \beta_{\rm l}  - \arcsin \beta_{\rm r} )/\pi.
\end{equation} 
We have plotted this ratio in Fig.~\ref{fraction} as function of the tilt angle.
Note that the value of $f$ only depends on the tilt angle and of course and
the ratio $p_{\rm h}/E$ of the hardness $p_{\rm h}$ and Young's modulus $E$.
One observes that $f$ is small for large tilt angles and substantial for small
tilt angles. 
With the value of $f$ one can map the problem for finite hardness on the purely
elastic theory (for infinite hardness). If $F_z$ is the force which pushes the skate
in the ice of hardness $p_{\rm h}$, then $F_{\rm el}= F_z/f$ is the force needed
to give the same indentation in a purely elastic medium. So computing the
indentation in the elastic limit suffices for the general case of a finite hardness.

\section{Conclusion}

The response of ice on a skate,  pushed with force $F_z=Mg$ into the ice, has been
exactly calculated in the elastic approximation. The skate intrudes deeper in the ice
with a growing tilt angle $\phi$. The upright skate has the smallest indentation.
As the pressure distribution diverges at the sharp edges, 
the linear elastic theory only applies in the limit of an infinite hardness $p_{\rm h}$.
However the hardness of ice
is relatively small, of the order of 10 MPa. The relevant dimensionless measure is the 
ratio of the hardness $p_{\rm h}$ and the effective Yong's modulus $E^*$, which is small
for ice: $p_{\rm h} /E^* \simeq 0.01$. Consequently the deformation is to a large extend
plastic and not elastic. We have introduced a fraction $f$, which is the ratio of external
force needed for an indentation at a finite hardness and the external force for the
same indentation in a pure elastic medium.
With $f$ we can map the plastic deformation under a load $F_z$ on
an purely elastic deformation under a load $F_{\rm el} = F_z/f$. Fortunately $f$ depends
only on the tilt angle of the skate.

The larger the tilt angle $\phi$, the smaller the fraction $f$ (see Fig.~\ref{fraction}).
The upright skate has the largest elastic fraction, but it is a separate case. The tilt angles
that we treat, have only one edge intruding into the ice. Due to the finite width
 $w=1.1$mm, the other edge will start to touch the ice for tilt angle $\phi \sim<0.01.$
We have not treated this intermediate region as we could not find the exact solution for
the deformation. The upright skate permits again an exact solution.

The tilted skate and the upright skate not only differ in geometry but also in
the way the free parameters appear in the solution. For the tilted skate the amplitude
$A$ is fixed by the slopes of the tilt and the freedom is in the width of the region
that makes contact with the ice. The upright skate has a fixed width $w$ of contact and
a free amplitude $B$ in the solution. In both cases the free parameter is determined by
the match between the asymptotic behaviour of the deformation of the finite skate and 
that of the analytic solution for a quasi infinite long skate.

The tilted skate can be continuously moved to the upright position and the tilted
solution should merge continuously into the upright solution. Coming back to the
three regimes of the introduction, we summarise the singularities of each.
\begin{itemize}
\item In the case of a sufficiently large tilt, one has the central logarithmic
  singularity in the pressure distribution at the edge. The slope of the surface of the
  ice makes a jump from $-1/\tan(\phi)$ to $\tan(\phi)$. At the points where the
  skate looses contact with the ice, the slope of the surface goes continuously from a
  prescribed value to a varying value. Also the pressure goes continuously to zero
  at these points.
\item In the small angle regime the pressure distribution has two logarithmic
  singularities at the edges of the skate blade. At one side the jump in the slope of
  the ice surface is from $-1/\tan(\phi)$ to $\tan(\phi)$ and on the other side it is
  a jump from the slope $\tan(\phi)$ underneath the skate to the finite slope of
  ice surface in contact with the air.
\item The perfect upright skate has two symmetric square root singularities
  in the pressure, accompanied by two jumps of the slope from zero underneath the
  skate to infinite outside the skate.
\end{itemize}
The first and last regime are exactly analysed and the middle regime is left open.

{\bf Acknowledgement} The author is indebted to Leen van Wijngaarden, Technical
University Twente, for stimulating discussions and useful suggestions, in particular
concerning the relevant literature.

\appendix

\section{The deformation of a long sharp wedge}\label{wedge}

In this section we discuss the function $H(y)$, which is the generator
of the functions to be used for the deformations. It has singular points at $y=0, \,
y=-y{\rm l}$ and $y=y_{\rm r}$ on the real axis and it is analytic for 
${\rm Im} \, y>0$. $H(y)$ reads
\begin{equation} \label{A1}
  H(y) = a_{\rm r} + \frac{A}{2 \pi i} \log \left( \frac{X(y)+\sqrt{Y(y)}}
    {X(y)-\sqrt{Y(y)}} \right),  \quad \quad 0<y<y_{\rm r},
\end{equation}
with the definitions
\begin{equation} \label{A2}
  A=a_{\rm r}+a_{\rm l}, \quad   X(y) = m + R_- y /2,
\quad   Y(y) =m^2 +R_- \, m y -y^2.
\end{equation}
This solution agrees with the one found for the symmetric wedge, for which $R_-=0$
\cite{john1}.

First we want to extend, by analytic continuation, the function $H(y)$ to the
other intervals. We note that $X(0)=\sqrt{Y(0)}=m $,
thus $H(y)$ has a (logarithmic) singularity at $y=0$
\begin{equation} \label{A3}
H(y) \simeq \frac{A}{2 \pi i} \log \frac{1}{y^2} + {\cal{O}} (1).
\end{equation} 
So moving from $y>0$ to $y<0$, the log picks up an imaginary part $-2\pi i$,
resulting in an extra contribution $-A$.
Therefore the function $H(y)$ is represented, in the interval $-y_{\rm l}<y<0$,  by
\begin{equation} \label{A4}
  H(y) = -a_{\rm l} +\frac{A}{2 \pi i} \log
  \left( \frac{X(y)+\sqrt{Y(y)}}{X(y)-\sqrt{Y(y)}} \right), \quad \quad -y_{\rm l}<y<0.
\end{equation}
Next we extend the function $H(y)$ to the interval $y > y_{\rm r} $. The function $Y(y)$
vanishes as a square root at $y=y_{\rm r}$. Continuing through the upper half $y$ plane
$\sqrt{y_{\rm r}-y}$ to $y>y_{\rm r}$ one picks up a factor $-i$ and $H(y)$ is given, for $y > y_{\rm r}  $ by
\begin{equation} \label{A5}
  H(y) = a_{\rm r} + \frac{A}{2 \pi i} \log
  \left( \frac{X(y)-i\sqrt{-Y(y)}}{X(y)+i\sqrt{-Y(y)}} \right), \quad \quad y > y_{\rm r}.
\end{equation}
We rewrite this using the relation
\begin{equation} \label{A6}
  \log \left( \frac{X-i\sqrt{-Y}}{X+i\sqrt{-Y}} \right) = -2i \arctan (\sqrt{-Y}/X),
\end{equation}
as
\begin{equation} \label{A7}
H(y) = a_{\rm r} - \frac{A}{\pi} \arctan (\sqrt{-Y(y)}/X(y)),  \quad \quad y > y_{\rm r}.
\end{equation}
The function $H(y)$ is in the last interval $y< -y_{\rm l}$ given by
\begin{equation} \label{A8}
  H(y) = -a_{\rm l} + \frac{A}{\pi }  \arctan (\sqrt{-Y(y)}/X(y)  \quad \quad y< -y_{\rm l}.
\end{equation}
using again Eq.~(\ref{A6}) and the fact that  the square root picks up a 
factor $i$ in passing from $y>-y_{\rm l} $ to $y < -y_{\rm l}$.

\subsection{Determination of $r$}
The ratio  $r$ is determined by the asymptotic behaviour of $H(y)$ for
large $y$. For the argument of the arctan in Eq.~(\ref{A7}) we find
\begin{equation} \label{A9}
  \frac{\sqrt{-Y(y \rightarrow \infty)}}{X(y \rightarrow \infty)} = 
\frac{2}{R_{-}}
\end{equation}
Thus $H(y)$ approaches asymptotically the value
\begin{equation} \label{A10}
  H(y \rightarrow \infty) =a_{\rm r} - \frac{A}{\pi} 
\arctan \left( \frac{2}{R_-} \right).
\end{equation}
In order that this value vanishes we have to require that
\begin{equation} \label{A11}
\tan \left( \frac{\pi a_{\rm r}}{A} \right)= \frac{2}{R_-}.
\end{equation} 
Introducing the angle $\chi$
\begin{equation} \label{A12}
\chi = \frac{\pi a_{\rm r}}{A}=\pi \sin^2 \phi, 
\end{equation} 
we may write Eq.~(\ref{A11}) also as
\begin{equation} \label{A13}
\frac{2}{R_-} = \tan \chi,
\end{equation} 
which is the condition for $r$, taking the form
\begin{equation} \label{A14}
r = \frac{1+ \cos(\chi)}{\sin(\chi)}= \frac{1}{\tan(\chi/2)}.
\end{equation}

We notice that, on the other side $y \rightarrow -\infty$, the  argument of the arctan
passes through infinity at the value $y=-2m/R_-$ where $X(y)=0$. This is not a
singular point of the original expression involving the log of $X(y) \pm \sqrt{Y(y)}$.
The value of the arctan approaches  $\pi/2$ at $X(y) \simeq 0$ and picks up an
additional contribution $\pi$ when $X(y) < 0$. So
\begin{equation} \label{A15}
  \arctan \left(\frac{\sqrt{-Y (y \rightarrow -\infty)}}{X(y \rightarrow -\infty)} \right)
  = \pi -\arctan \left( \frac{2}{R_-} \right).
\end{equation}
Using  Eq.~(\ref{A12}) to work out the arctan we get to fulfil 
\begin{equation} \label{A16}
0 = -a_{\rm l} +\frac{A}{\pi}  \left( \pi - \frac{\pi a_{\rm r}}{A} \right),
\end{equation}
which is an identity. So the value for $r$ as following from Eq.~(\ref{A12})
guarantees also the proper asymptotic behaviour at $s \rightarrow -\infty$.
A convenient relation is
\begin{equation} \label{A17}
  \frac{R_-}{R_+} = \cos \chi.
\end{equation} 

\subsection{Properties of $H(y)$}

In view of the integrations for the pressure and the shape of the surface it is
useful to note that the derivative of the function $H(y)$ is relatively simple.
We find inside the basin
\begin{equation} \label{A18}
  \frac{d H(y)}{dy} 
  = \frac{A m i}{\pi y \sqrt{Y(y)}} \quad \quad -y_{\rm l} < y < y_{\rm r}.
\end{equation}
We can  use this relation for the integral over $p(y)$.  
We find by partial integration the primitive of $p(y)$
\begin{equation} \label{A20}
  \int  p(y)dy = y p(y) + \frac{A m}{\pi} \int \frac{dy}{\sqrt{Y(y)}} =
  y p(y) + \frac{A m}{\pi} \arcsin \left(\frac{2 y -y_{\rm r} +y_{\rm l}}{y_{\rm r} +y_{\rm l}}) \right).
\end{equation}
Applying this result to the whole interval $-y_{\rm l} < y < y_{\rm r}$ gives
\begin{equation} \label{A21}
  \int^{y_{\rm r}}_{-y_{\rm l}}  dy \, p(y) =  \frac{Am}{\pi} (\arcsin 1 - \arcsin( - 1)) = A m.
\end{equation}

Likewise we find outside the basin 
\begin{equation} \label{A22}
  \frac{d H(y)}{dy} = - \frac{A m}{\pi y \sqrt{-Y(y)}}, \quad \quad  y> y_{\rm r}.
\end{equation}
for the asymptotic behaviour of $H(y)$. The integral of the real part of $H(y)$ is
\begin{equation} \label{A23}
  \int H(y)dy = y H(y) +\frac{A}{\pi}  \int \frac{dy}{\sqrt{Y(-y)}} =
  y H(y) + \frac{Am}{\pi}  {\rm acosh} \left( \frac{2 y -y_{\rm r} +y_{\rm l}}{y_{\rm r} +y_{\rm l}}) \right).
\end{equation}
The primitive of $H(y)$  determines the asymptotic behaviour of $q(y)$ for large $y$.
  
In addition we need the asymptotic behaviour of $H(y)$ itself. For large $y$ we find
\begin{equation} \label{A24}
  \frac{\sqrt{-Y(y)}}{X(y)}= \frac{2}{R_-} - \frac{R^2_+ m}{R^2_- y} 
    + \cdots .
\end{equation}
Expanding the arctan around the value $2/R_-$ gives
\begin{equation} \label{A25}
  \arctan \frac{\sqrt{-Y(y)}}{X(y)} = \arctan  \frac{2}{R_-} - \frac{m}{y} + \cdots
\end{equation} 
Thus the asymptotic behaviour of $H(y)$ is given by 
\begin{equation} \label{A26}
  H(y) =a_r -\frac{A}{\pi} \left(\arctan \frac{2}{R_-} - \frac{m}{y} + \cdots \right),
\end{equation} 
or in view of Eq.~(\ref{A10}) given by
\begin{equation} \label{A27}
H(y \rightarrow \infty) \simeq  \frac{A m}{\pi y}.
\end{equation}

\section{The analytic solution for the upright skate} \label{upr}

For the upright skate the function $H(y)$ has the form \cite{musk}
\begin{equation} \label{B1}
    H(y) = \frac{B}{i\sqrt{w^2/4-y^2}} \quad \quad \quad -w/2 < y < w/2, 
\end{equation}
with $B$ a positive real constant. Outside the basin one finds $H(y)$ by
analytic continuation through the upper half complex $s$ plane.
From Eq.~(\ref{B1}) we find for the pressure underneath the skate
\begin{equation} \label{B2}
  p(y) = \frac{B}{\sqrt{w^2/4-y^2}} \quad \quad \quad -w/2 < y < w/2.
\end{equation}
Outside the basin the function $H(y)$ becomes real and $p(y)=0$. At the
edges the pressure diverges as a square root. So the pressure will exceed
the hardness at the edges and the deformation causing the deformation to 
become plastic. 
Inside the basin $q(y)$ vanishes since the skate is flat and upright.
Outside the basin the function $q(y)$ becomes
\begin{equation} \label{B3}
  q(y) = \frac{B}{\sqrt{y^2-w^2/4}}, \quad \quad y <-w/2 \quad
  {\rm and} \quad  y > w/2.
\end{equation} 
The amplitude $B(x)$ is function of $x$ and  follows from a match between 
the asymptotic decay of the deformation of a finite skate and the 
exact solution in the region $w \ll y \ll l$.
The expression for the asymptotic decay of a finite skate is as in Eq.~(\ref{f5})
\begin{equation} \label{B4}
  u_z(x,y)  \simeq  -  \frac{1 }{2 \pi } \int^l_{-l} dx' 
\int^{w/2}_{-w/2} dy' \frac{p(y')}{[(x'-x)^2+(y-y')^2]^{1/2}}.
\end{equation}
For $y \gg w$ we may drop the $y'$ dependence in the denominator and carry
out the integration over $y'$
\begin{equation} \label{B5}
 \int^{w/2}_{-w/2} dy' p(y') = \pi B(x).
\end{equation}
Then the integral over $x'$ is elementary with the result
\begin{equation} \label{B6}
u_z (x,y) = -\frac{B(x)}{2} \left[ {\rm asinh} ((l-x)/y) - {\rm asinh} ((-l-x)/y) \right],
\end{equation} 
which behaves asymptotically as
\begin{equation} \label{B7}
u_x (x,y)  \simeq - \frac{B(x)}{2} \log \left( \frac{4(l^2-x^2)}{y^2} \right).
\end{equation} 
This behaviour has to be compared with the exact solution 
\begin{equation} \label{B8}
  u_z (x,y) = -d(x) + \int^y_{w/2} dy \, q(y) = -d(x) + \frac{B(x)}{2}\log
  \left(\frac{y +\sqrt{y^2-w^2/4}}{y - \sqrt{y^2-w^2/4}} \right),
\end{equation}
yielding asymptotically for the surface deformation
\begin{equation} \label{B9}
  u_z(x,y) \simeq -d(x)- B(x)\log (4y/w).
\end{equation}
Comparing Eqns.~(\ref{B7}) and (\ref{B9}) we see the the amplitudes of the 
$\log(y)$ terms are equal. The two forms coincide if also the constant terms
agree which is the case when
\begin{equation} \label{B10}
  B(x) = \frac{d(x)}{ \log(8(\sqrt{l^2-x^2}/w }).
\end{equation}
 This relation is the equivalent of Eq.~(\ref{f11}) for the tilted skate.
 
\section{The elastic equations}\label{elas}

The basic equations of linear elastic deformation are given as \cite{landau}
\begin{equation} \label{b1}
(1 - 2 \nu) \Delta {\bf u} + \nabla (\nabla \cdot {\bf u}) = 0. 
\end{equation}
where $\bf u$ is the deformation field and $\nu$ is Poisson's ratio.
The deformation generates a reaction force distribution $\bf f$ at
the surface, which is given by
\begin{equation} \label{b2}
g {\bf f}= [(1 -2 \nu) ({\bf n} \cdot \nabla {\bf u} + 
\nabla ({\bf n} \cdot {\bf u})) + 2 \nu \, {\bf n} ( \nabla \cdot {\bf u}),
\end{equation} 
with $g$ equal to
\begin{equation} \label{b3}
g = 2(1+\nu)(1-2\nu)/E
\end{equation}
and $E$ Young's modulus of  ice $E=0.88 \cdot 10^9$ Pa.
$\bf n$ is the normal to the surface.

Due to the large curvature radius $R$ the variations in the $x$ are slow and
we ignore them in the elastic equations. This reduces the elastic equations to the
plane theory of deformation, which read
\begin{equation} \label{b4}
\left\{ \begin{array}{rcl}
2(1-\nu) u_{y,yy} + (1-2 \nu) u_{y,zz} + u_{z,zy} & = & 0, \\*[3mm]
2(1-\nu) u_{z,zz} + (1-2 \nu) u_{z,yy} + u_{y,yz} & = & 0,
\end{array} \right.
\end{equation}
where subscripts after the comma refer to differentiations with respect to $y$ and $z$.
Likewise we find the force distribution as
\begin{equation} \label{b5}
\left\{ \begin{array}{rcl}
          g f_y & =  & [2(1-\nu)u_{y,y} +2 \nu u_{z,z}] n_y + (1-2\nu)(u_{y,z}+u_{z,y}) n_z
          \\*[3mm]
          g f_z & = & (1-2\nu)(u_{y,z}+u_{z,y}) n_y + [2(1-\nu) u_{z,z} + 2 \nu u_{y,y} ] n_z 
\end{array} \right.
\end{equation} 
More physical are the normal and tangential force, $f_{\rm n}$ and $f_{\rm t}$
defined as 
\begin{equation} \label{b6}
f_{\rm n} = n_y f_y  + n_z f_z, \quad \quad \quad f_{\rm t} = n_y f_z -n_z f_y.
\end{equation} 
Using the Eqns.~(\ref{b5}) we find for them the expressions
\begin{equation} \label{ b7}
\left\{ \begin{array}{rcl}
          g f_{\rm t} & = & (1-2 \nu) [ (2 n_y^2 -1)(u_{y,z}+u_{z,y}) +
                            2 n_y n_z (u_{z,z}-u_{y,y})]\\*[3mm]
          g f_{\rm n} & = & 2(1-\nu) u_{z,z} + 2 \nu u_{y,y} +
                            2(1-2 \nu) [n^2_y (u_{y,y} - u_{z,z}) + 2  n_y n_z (u_{y,z}+u_{z,y})]
\end{array} \right.
\end{equation} 

Since skates are polished the tangential force vanishes, not only outside the basin,
but also inside, i.e. everywhere. Using $f_{\rm t}=0$  the equation for $f_{\rm n}$
becomes
\begin{equation} \label{b8}
g f_{\rm n} = 2(1-\nu) u_{z,z} + 2 \nu u_{y,y} + 2(1-2 \nu) \frac{ n^2_y}{1 - 2 n^2_y} 
(u_{y,z}+ u_{z,y}).
\end{equation} 
It can be shown (see Appendix \ref{biharm}) that these equations can be solved 
by setting
\begin{equation} \label{b9}
u_{y,y}=u_{z,z}, \quad \quad \quad y_{y,z}+y_{z,y}=0.
\end{equation} 
The second equality follows from the first and the requirement that $f_{\rm t}=0$.
The normal force then obtains the form
\begin{equation} \label{b10}
g f_{\rm n} = 2 u_{z,z}
\end{equation} 
The pressure $p(y)$ is proportional to the normal force 
\begin{equation} \label{b11} 
p(y)= \frac{2}{E^*} f_{\rm n} (y) = 2 \frac{1-\nu}{1-2 \nu} u_{z,z}
\end{equation} 
The plain theory of deformation implies that 
\begin{equation} \label{b12}
q(y) = u_{z,y}
\end{equation} 
and $p(y)$ are real and imaginary parts of a complex function $H(y)$, 
which is analytic in the upper half $y$ plane.

\section{Expansion in biharmonic functions}\label{biharm}

We  restrict ourselves to the case where there is mirror symmetry in the
$y$ direction, as is the case in the upright skate. Then
$u_y (y,z)$ is an odd function of $y$ and $u_z(y,z)$ and even function of $y$.
The general case runs similar with an expansion of the even and odd parts.
The  following expansion in biharmonic functions then suffices.
\begin{equation} \label{C1}
\left\{ \begin{array}{rcl}
          u_y (y, z) & = & \displaystyle \int^\infty_0 dk \, [\alpha(k) + \alpha'(k) z]
                         \exp(-kz) \sin(ky), \\[4mm]
          u_z (y, z) & = &  \displaystyle \int^\infty_0 dk \, [\gamma(k) + \gamma'(k)z]
                         \exp(-kz) \cos(ky).
\end{array} \right.
\end{equation}
Note that the first bulk equation  involves only odd terms such that
only the terms with $\sin(ky)$ appear.
Inserting the expansion Eq.~(\ref{C1}) into the first bulk equation 
leads to the equation for the coefficients
\begin{equation} \label{C2}
  \begin{array}{l}
    2(1-\nu) [\alpha(k)+\alpha'(k) z] k^2 = \\*[2mm]
    (1- 2 \nu) [(\alpha(k) + \alpha'(k) z) k^2
    - 2 \alpha'(k) k] + (\gamma(k) + \gamma'(k) z )k^2 -\gamma'(k) k.
   \end{array}
\end{equation}  
The terms with and without $z$ have to match and that implies the relations
\begin{equation} \label{C3}
\gamma'(k) =\alpha'(k), \quad \quad (3-4 \nu) \alpha'(k) =k (\gamma(k) - \alpha(k)).
\end{equation} 
The second bulk  involves only even functions in $y$ and 
therefore only terms with $\cos(ky)$ appear. We then have to fulfil again the 
relation between the coefficients
\begin{equation} \label{C4}
\begin{array}{l}
  2(1-\nu)[k^2 (\gamma(k) + \gamma'(k) z )- k \gamma'(k) ] =  \\*[2mm]
  (1 -2 \nu) [\gamma(k) +
  k^2  \gamma'(k) z ] + k^2 (\alpha(k)+ \alpha'(k) z ) - k \alpha'(k).
 \end{array}                                                                        
\end{equation} 
Surprisingly enough equating the terms with and without $z$ leads to exactly
the same relations Eq.~(\ref{C4}). This means that we can fulfil the bulk
equations with the expansion (\ref{C1}) and still have 
$\alpha(k)$ and $\gamma(k)$ as free coefficients.

So we eliminate $\alpha'(k)$ and $\gamma'(k)$ with Eq.~(\ref{C3}) and
get for the relevant functions at the surface $z=0$
\begin{equation} \label{C5}
\left\{ \begin{array}{rcl}
u_{y,y} (y,0) & = &\displaystyle \int^\infty_0 dk \,  k \, \alpha(k) \cos (k y), \\*[4mm]
u_{y,z} (y,0) & = &\displaystyle  \frac{1}{3-4 \nu} \int^\infty_0 dk \, k\,
                          [ - 4(1-\nu) \alpha(k) + \gamma(k) ] \sin (k y), \\*[4mm]
u_{z,y} (y,0) & = & \displaystyle -\int^\infty_0 dk \, k \, \gamma(k) \sin (k y), \\*[4mm]
u_{z,z} (y,0) & =& \displaystyle \frac{-1}{3-4 \nu} \displaystyle \int^\infty_0 dk \, k\,
                         [ \alpha(k) + 2(1-2\nu) \gamma(k)] \cos(k y).
\end{array} \right.
\end{equation}

We now note that if we take
\begin{equation} \label{C6}
\alpha(k) = -\frac{1 - 2\nu}{2(1-\nu)} \gamma(k),
\end{equation}
The following combinations  vanish
\begin{equation} \label{C7}
u_{y,z} (y,0) + u_{z,y} (y,0) =0 \quad \quad {\rm and} \quad \quad 
u_{z,z}(y,0) -u_{y,y} (y,0) = 0.
\end{equation} 
Inserting Eq.~(\ref{C6}) into the expressions (\ref{C5}) for the functions $u_z$ and $u_y$,
yield
\begin{equation} \label{C8}
u_{z,z} (y,0) = u_{y,y} (y,0) = - \frac{1-2 \nu}{2(1-\nu)} 
\int^\infty_0 dk \,  k \, \gamma (k) \cos (k y).
\end{equation} 
This means that in Eq.~(\ref{b9}) one has indeed $f_{\rm t} =0$ and that 
the expression for $f_{\rm n}$ reduces to Eq.~(\ref{b10}).
In Eq.~(\ref{b9}) the normal force is related to the function $p(y)$ for which we now 
find the expansion
\begin{equation} \label{C9}
  p(y) =-\int^{\infty}_0 dk \, k \, \gamma (k) \cos (ky).
\end{equation}

We see from the definition in Eq.~(\ref{b6}) that the function $q(y)$
is given by
\begin{equation} \label{C10}
  q(y) = -\int^{\infty}_0 dk \, k \, \gamma (k) \sin (ky).
\end{equation}   
Thus $q(y)$ and $p(y)$ are real and imaginary parts of the function
\begin{equation} \label{C11}
  H(y) = i \int^{\infty}_0 dk \, k \, \gamma (k) \exp (iky).
\end{equation}   
which is analytic in the upper half complex $y$ plane. 

In order to calculate the complete deformation we have to determine the
expansion coefficients $\gamma (k)$ from the inverse fourier relation
\begin{equation} \label{C12}
  k \gamma (k) = \frac{1}{2} \int^{w/2}_{-w/2} p(y) \cos (ky), 
\end{equation}
with $p(y)$ given by Eq.~(\ref{B2}). With the value of the $\gamma (k) $
one can calculate all the other expansion coefficient and thus determine the
complete deformation field.

\end{document}